\documentstyle[12pt]{article}

\begin{document}

\topmargin 0pt
\oddsidemargin 5mm

\setcounter{page}{1}
\begin{titlepage}
\vspace{2cm}
\begin{center}

{\bf System of 2 Derrida's models with fixed correlation between the
spin configurations of subsystems}\\
\vspace{5mm}
{\large 
Allakhverdian A., Saakian D.}\\
\vspace{5mm}
{\em Yerevan Physics Institute}\\
{Alikhanian Brothers St.2, Yerevan 375036, Armenia\\
and
Laboratory of Computing techniques and automation, JINR,
141980 Dubna, Russia\\
Saakian @ vx1.YERPHI.AM
\\
PACS.05.50+q-Lattice theory and statics, Ising problems\\
PACS.75.10Nr-Spin-Glass and other random models}
\end{center}

\vspace{5mm}
\centerline{\bf{Abstract}}

 The free energy and finite size effects are
calculated for the system of 2 Derrida's models with fixed constraint between
 the spin configurations. Calculation were performed for the Random Energy
Model approach.
\end{titlepage}

%\section{Introduction}
\vspace{5mm}

The random energy model (REM) was considered in \cite{BD}, where
it has been derived from the hamiltonian of random $P $-spin
interactions. It has been solved directly in [1]. Then Gross and Mezard [2]
solved Derrida's P-spin model by means of mean-field method (Parisi ansatz
for replica symmetry breaking). This 2 models are the same in the thermodynamics
limit. Perhaps accuracy is even better, exponential for the good choice of P
(P=N/2) [3].\\
It is possible to use Derrida's model for the optimal coding [4-7]. This Idea
of Sourlas was applied to the ferromagnetic phase of Derrida's like model,
which gives optimal coding for the simple symmetric channel.\\
To consider simple case of multi-terminal systems (the case of correlated
sources) we need to consider system of 2 models of Derrida's, with fixed
correlation among their spin configurations. In this work we are going to
give REM version of such model.\\

Let's consider 2 sets of $2^N$ energy levels $E_\alpha ^1,E_\alpha ^2
~ 1\leq\alpha\leq 2^N; E=E_\alpha ^1+ E_\beta ^2$. In addition not all the
pairs $E_\alpha ^1, E_\beta  ^2$ are permitted.

The matrix of connectivity $'_{\alpha \beta  }$ was introduced,
which elements takes the value 1 in the case when the
pair $(E_\alpha , E_\beta) $ and $0$-otherwise.

The energy levels $E^1_\alpha  $ have the distribution:
\begin{equation}
\label{AA}
P_x(E_\alpha  ^1)=\frac{1}{\sqrt{\pi N}}\,{\rm exp}
\left[-(E_\alpha^ 1)^2/N\right]
\end{equation}
and for $E_\alpha  ^2$ the distribution is:
\begin{equation}
\label{AB}
P_y(E_\beta ^2)=\frac{1}{\sqrt{\pi N}J}\,{\rm exp}
\left[-(E_\beta^ 2)^2/NJ^2\right]
\end{equation}
For the free energy we have an expression
\begin{equation}
\label{AC}
<\ln Z>=\int\,\prod^{M}_{\alpha  =1} dE_\alpha  ^1 dE_\alpha ^2
\ln\left[ \sum^{M }_{\alpha =1\atop \beta=1}\, C_{\alpha  \beta}
{\rm exp}\left[E_\alpha^1+E_\beta^2 \right]B \right]
P_x(E_\alpha ^1)P_y(E_\beta ^2);\ M=2^N
\end{equation}
There is a condition for the matrix $C_{\alpha \beta}$
\begin{equation}
\label{AD}
\sum_{\beta =1}^{M}C_{\alpha \beta }=e^{N\ln 2h};\quad\alpha
=1,\dots,M.
\end{equation}
By using the trick from \cite{BD}
\begin{equation}
\label{AE}
\ln Z=\Gamma '(1)-\int\limits_{o}^{\infty }\ln
Zde^{-tz}\equiv \Gamma '(1)-\int\limits_{-\infty }^{\infty
}ude^{-e^uZ}
\end{equation}
one gets $(\lambda _1=N\sqrt{\beta };\quad \lambda
_2=N\sqrt{\beta }J)$
\begin{equation}
\label{AF}
<\ln Z>=\Gamma '(1)-\int\limits_{-\infty }^{\infty }ud\Psi(u)
\end{equation}
where
\begin{equation}
\label{AG}
\Psi (u)=\int\limits_{-\infty }^{\infty }\prod_{\alpha }\frac{dx_\alpha
dy_\alpha }{\pi }e^{-\sum_{\alpha }(x^2+y^2)-e^u\sum_{\alpha
,\beta }C_{\alpha \beta }e^{\lambda _1x_\alpha+\lambda _2y_\beta }}
\end{equation}
In the bulk approximation our monotonic function $\Psi (u)$ is similar to  step function
(with the center in the point $u_0$, will be defined later) and one derives
\begin{equation}
\label{AH}
\ln Z\simeq -u_0
\end{equation}
For $u>-u_0$ the function $\Psi(u) $ decreases exponentially, and
for  $u<-u_0$ it takes 1 with exponential precision. Depending on
the values  of $\lambda _1,\lambda _2$ we obtain different
asymptotics for the function:
\begin{equation}
\label{AJ}
\Psi_1 (u)=\int\limits_{-\infty }^{+\infty }\frac{dx
dy}{\pi }e^{-(x^2+y^2)-e^u(e^{\lambda _1x+\lambda _2y})}
\end{equation}
In the case $|u|\rightarrow \infty $ we have 2
possiblities.\\
At first  we can expand the last term in the exponent.
It yields:
\begin{equation}
\label{AK}
\Psi_1 (u)\simeq 1-e^{u+(\lambda _1^2+\lambda _2^2)/4}
\end{equation}
The main contribution (\ref{AK}) comes from the domain near to
the point:
\begin{equation}
\label{AL}
x=\frac{\lambda _1}{2};\quad \quad y=\frac{\lambda _2}{2}
\end{equation}
For the calculation to be self-consistent, ${\rm
exp}\left[u+(\lambda_1x+\lambda _2y) \right]$ must be small
in the point (\ref{AL}).
Hence one gets the condition:
\begin{equation}
\label{AW}
u<-(\lambda _1^2+\lambda _2^2)/2
\end{equation}
Doing the same for all the degrees of freedom, we  get:
\begin{equation}
\label{AZ}
\Psi\left(u\right)\simeq
\exp\left[-e^{u+\frac{\lambda_1^2+\lambda_2^2}{4}+N\ln2(1+h)}\right]
\end{equation}

Hence as in (\ref{AE}) we obtain:
\begin{equation}
\label{AR}
\frac{\ln Z}{N}=\frac{\lambda _1^2+\lambda _2^2}{4}+\ln
2(1+h)\equiv(1+J^2)\frac{B ^2}{4}+\ln 2 (1+h)
\end{equation}
The condition (\ref{AW})yields:
\begin{equation}
\label{RR}
B  < B _c\equiv2\sqrt{\frac{(1+h)\ln 2}{1+J^2}}
\end{equation}
Now let it be the situation, when $e^{u+\lambda _1x+\lambda _2y}$
is big  in the region, giving the main contribution to the
integral. Then:
\begin{equation}
\label{AT}
\Psi (u)=\int\limits_{-\infty }\prod_{\alpha=1 }^{M} d x_\alpha
d y_\alpha  e^{-\sum_{\alpha }(x^2_\alpha +y^2_\alpha )}
\end{equation}
where the convex domain $D$ is obtained from the condition:
\begin{equation}
\label{AY}
\lambda _1x_\alpha + \lambda _2y_\beta < -u
\end{equation}
The value of the integral (\ref{AT}) will be defined by the point
on the boundary $D$, where $(x_\alpha ^2+y^2_\alpha )$ is
minimal.

For that point we get:
\begin{equation}
\label{AU}
x_c=-\frac{\lambda _1 u}{\lambda _1^2+\lambda _2^2};\quad \quad
y_c=-\frac{\lambda _2 u}{\lambda _1^2+\lambda _2^2}
\end{equation}
For one pair it'll be derived (omitted pre exponents):
\begin{equation}
\label{AI}
\Psi_1 (u)\simeq 1-e^{-(x_c^2+y_c^2)}
\end{equation}
The same goes with $\Psi (u)$:

\begin{equation}
\label{AO}
\Psi (u)\simeq {\rm exp}\left[-e^{\frac{-u^2}{\lambda _1^2+\lambda
_2^2}+N \ln  2(1+h)}\right]
\end{equation}
Hence we get (for $B>B_c $):
\begin{equation}
\label{BA}
\frac{\ln Z}{N}\simeq \sqrt{(\lambda _1^2+\lambda _2^2)N(1+h)\ln 2}\equiv BN
\sqrt{(1+h)(1+y^2)\ln2}
\end{equation}
For the general case for $\Psi_1(u) $  we have the expansion:
\begin{equation}
\label{BB}
\Psi_1(u)\approx \int\limits_{-\infty }^{+\infty
}dx\frac{e^{-x^2}}{\sqrt{\pi }}\left\{1-\theta (-u-\lambda
_1x-\frac{\lambda _2^2}{2}) e^{u+\lambda _1+\lambda
_2^2/4}-\theta (u+\lambda _1x+\frac{\lambda
_2^2}{2})e^{-(u+\lambda _1x)^2/\lambda _2^2}\right\}
\end{equation}
By taking the integral for one $x_\alpha $ and $2^{Nh}$of $y_\beta$
 connected with it, we get the $2^{Nh}$ power of the
expression in the braces in (of) (\ref{BB}). This yields to the truncation
of the integration domain in (\ref{BB}).
Depending on$\lambda _1, \lambda _2$ we get:
\begin{equation}
\label{BC}
\frac{1}{\sqrt{\pi }}\,\int\limits_{-\infty
}^{x_1}e^{-x^2}dx \simeq 1-\frac{e^{-x_1^2}}{2x_1\sqrt{\pi }};\quad
x_1=\left(\frac{\lambda _2^2}{4}+Nh\ln2+u \right)\frac{1}{\lambda
_1}
\end{equation}
or
\begin{equation}
\label{BD}
\frac{1}{\sqrt{\pi }}\,\int\limits_{-\infty
}^{x_2}e^{-x^2}dx \simeq 1-\frac{e^{-x_2^2}}{2\sqrt{\pi}x_2};\quad
x_2=\frac{\lambda _2\sqrt{Nh\ln2}+u}{\lambda_1}
\end{equation}
If we raise (\ref{BC}), (\ref{BD}) to the $2^N$ power we obtain
the expression for $\Psi(u)$:
\begin{equation}
\label{BE}
\Psi(u)\simeq{\rm exp}\left[-e^{\frac
{-(u+Nh\ln 2+\lambda _2^2/4)}{\lambda _1^2}+N\ln 2}\right]
\end{equation}
or
\begin{equation}
\label{BR}
\Psi(u)\simeq{\rm exp}\left[-e^{
-(u+\lambda _2 \sqrt{Nh\ln 2}/\lambda _1^2+N\ln 2}\right]
\end{equation}
Hence we find, that for $2\sqrt{\ln 2}\,<\,B<\,2\sqrt{h \ln 2}/J$
\begin{equation}
\label{BF}
\frac{\ln z}{N}=\lambda _1\sqrt{Nh\ln 2}+\frac{\lambda
_1^2}{4}+h\ln 2\equiv\beta \sqrt{\ln 2}+\frac{\beta ^2J^2}{4}
+h\ln 2
\end{equation}
and for $\beta \,>\,2\sqrt{h \ln 2}/J$
\begin{equation}
\label{BG}
\frac{\ln z}{N}=x_1\sqrt{N\ln 2}+\lambda
_2\sqrt{h\ln 2}\equiv\beta \sqrt{\ln 2}\frac{\beta ^2y^2}{4} +h\ln 2
\end{equation}
The solutions (\ref{BF}),(\ref{BG}) are correct for
\begin{equation}
\label{BH}
h\,>\,J^2
\end{equation}
and for
\begin{equation}
\label{BJ}
h\,<\,J^2
\end{equation}
the solution is (\ref{BA})
At high temperatures we have paramagnetic solution (14). Then, in
the case (30) system falls into SG phase (21) at $B=B_c$.\\
 In the case (29) we have a mixture phase (SG+paramagnetic)(27) and SG-(28).
Now let's consider the ferromagnetic variant of the models.

For $E_1^1, E_1^2$ we have (instead of (\ref{AA})) and (\ref{AB}):

\begin{equation}
\label{BK}
P_1(E_1^1)=\frac{1}{\sqrt{\pi  N}}{\rm exp}\left[-(E_1^2+J_0^1
N)^2/N\right]
\end{equation}
\begin{equation}
\label{BL}
P_2(E_1^2)=\frac{1}{\sqrt{\pi  N}J}{\rm exp}\left[-(E_1^2+J_0^2
N)^2/J^2 N\right]
\end{equation}
For other $E_\alpha ^1, E_\alpha ^2;\alpha \geq 2 $ we have the
old distributions (\ref{AA}),(\ref{AB}).

The constants $J_0 ^1,$ and $J_0 ^2$ are positive.

In this case for $\Psi(u)$ we obtain:

\begin{eqnarray}
\label{CB}
\Psi(u)&=&\int\prod_{\alpha}\frac{dx_\alpha dy_\alpha}{\sqrt{\pi}}\exp\left\{-\sum_{\alpha }(x^2_\alpha +y^2_\alpha )-\left\{
{\rm exp}\left[ u+u_1+u_2+\lambda _1x_1+\lambda _2y_2\right]-
\right.\right.\nonumber\\
&&\nonumber\\
&&-\sum_{\beta }C_{1\beta }{\rm exp}\left[ u+ u_1+\lambda
_1x+\lambda _2 y\right]
-\sum_{\alpha }C_{\alpha 2 }{\rm exp}
\left[ u+ u_2+\lambda_1x+\lambda _2 y\right]-\nonumber\\
&&\left.-\sum_{\alpha ,\beta
=2}^{M}C_{2\beta }{\rm exp}\left[ u+ \lambda_1x+\lambda _2 y\right]
\}\right\}
\end{eqnarray}
where
\begin{equation}
\label{BV}
u_1=J_0^1 B N\quad \quad u_2=J_0^2 B N
\end{equation}
For the ferromagnetic phase the first term in braces in
(\ref{CB}) is resolving. In the bulk approximation the three others can be
neglected, and then $\Psi(u)$ behaves like a step-function in the
point -$ (J_0^1+J_0^2)N_ B$, to the right from this point it damps like:

\begin{equation}
\label{CL}
\Psi(u)\simeq{\rm exp}\left[ -\frac{u^2}{\lambda _1^2+\lambda
_2^2}\right]
\end{equation}
Hence we come to:
\begin{equation}
\label{BX}
<\frac{\ln Z}{N}>=(J_0^1+J_0^2)N_ B
\end{equation}
This solution is correct, when it's bigger than the corresponding
expression for the spin-glass and mixed (one system is in
ferromagnetic phase, the other - in the spin - glass). Mixed
phase is derived, when in  braces in (\ref{CB}) the second and
the third terms are dominant. Then we get:
\begin{equation}
\label{DA}
\frac{\ln Z}{N}=J_0^1 N_ B+B \sqrt{h\ln 2}J
\end{equation}
or
\begin{equation}
\label{DB}
\frac{<\ln Z>}{N}=J_0^2 N_ B+B \sqrt{h\ln 2}
\end{equation}
In the case (\ref{BH}) we get the following condition for the
existence of the solution (\ref{BX})

\begin{equation} \label{A39} \left\{
\begin{array}{l} J_0^1+J_0^2\,>\,\sqrt{\ln2}+\sqrt{h\ln2}J\\
J_0^1\,>\,\sqrt{h\ln2}\\ J_0^2\,>\,\sqrt{h\ln2}J\\
\end{array}\right.
\end{equation}
If we have the variant (\ref{BJ}) for existence of (\ref{BX}) we
get:
\begin{equation} \label{A40} \left\{
\begin{array}{l} J_0^1+J_0^2\,>\,\sqrt{\ln2(1+h)(1+J^2)}\\
J_0^1\,>\,\sqrt{h\ln2}\\
J_0^2\,>\,\sqrt{h\ln2}J\\
\end{array}\right.
\end{equation}

Let's consider now the finite size effects. For the calculation
of the magnetization we come to the expression:
\begin{equation}
\label{BM}
\int\limits_{-\infty}^{+\infty} \Psi_2(u)du
\end{equation}
\begin{eqnarray}
\label{BN}
&&\Psi_2(u)=\int\prod_{\alpha}\frac{dx_\alpha dy_\alpha }{\pi}
\exp\left\{-\sum_{\alpha }(x^2_\alpha +y^2_\alpha)+u+u_1+u_2+\lambda
_1x_1+\lambda _2y_1-\right.\nonumber\\
&&\nonumber\\
&&\qquad-\{
{\rm exp}\left[ u+u_1+u_2+\lambda _1x_1+\lambda _2y_1\right]+
\sum_{\beta }C_{1\beta}{\rm exp}\left[ u+ u_1+\lambda
_1x_1+\lambda_2 y_\beta\right]+\\
&&\left.\qquad+\sum_{\alpha }C_{\alpha 2}{\rm exp}\left[ u+ u_2+\lambda
_1 x_\alpha+\lambda _2 y_1\right]+\sum_{\alpha,\beta}C_{\alpha \beta}{\rm exp}
\left[u+ \lambda_1x_2+\lambda _2 y_\beta \right]\}\right\}\nonumber
\end{eqnarray}
If we neglect the three last terms in braces in (\ref{BL}) the
integral in (\ref{BN}) gives 1.

 The correction terms in the exponent leads to that (\ref{BN})
 is substituted by the expression like:
 \begin{equation}
\label{XZ}
\int\limits_{-\infty}^{-u_c} \Psi_2(u)\Psi_3(u)
\end{equation}
where $-u_c$ is a maximum for (\ref{DA}), (\ref{DB}),(\ref{BA})
in the case $h>J^2$ or for (\ref{DA}), (\ref{DB}), (\ref{BG})
in the case $h<J^2$. The function $\Psi_3(u)$ is defined by the
corresponding term in (\ref{BN}). It is like a step function in
the point $u_c$. When the dominant is (\ref{BG}) we get:
\begin{eqnarray}
\label{XB}
&&\int\limits_{-\infty}^{-u_c} \Psi_2(u)\left\{1-
\exp\left[ -\frac{u^2}{N\beta ^2(1+y^2)}+N\ln 2(1+h)
\right]\right\}\simeq\nonumber \\
&&\simeq 1-\int\limits^{-\infty}_{-u_c} \Psi_2(u)
-\int\limits_{-\infty}^{-u_c}
\Psi_2(u)\exp\left[ -\frac{u^2}{N\beta ^2(1+y^2)}+N\ln 2(1+h)
\right]
\end{eqnarray}
For the function $\Psi_2(u)$: it's equals to unity in the
vicinity of the point $-(J_0^1+J_0^2)/N B $, further it falls
exponentially.
\begin{equation}
\label{XC}
\Psi_2(u)\simeq \frac{1}{\sqrt{\pi } }\,\frac{N\beta
^2(1+y^2)}{2^{|u+(J_0^1+J_0^2)N\beta
|}}\exp\left\{-\frac{u+(J_0^1+J_0^2)N\beta ^2}{N\beta ^2(1+y^2)} \right\}
\end{equation}
The expression (\ref{XB}) can be substituted by (\ref{XB}) lies
when the asymptotic (\ref{XC}) is correct.

At last we obtain for the magnetization:
\begin{eqnarray}
\label{XN}
&&1-a\exp\left\{ -\frac{J_0^1+J_0^2+\sqrt{(1+J^2)(m+h)\ln
2}}{1+J^2}\,N\right\}
\theta
\left[-\frac{J_0^1+J_0^2}{\sqrt{1+J^2}}+2\,\sqrt{(1+h)\ln
2}\right]-\nonumber \\
&&-b\exp\left\{-\frac{(J_0^1+J_0^2)^2N}{1+J^2}+N\ln2(1+h)\right\}
\theta \left( \frac{J_0^1+J_0^2}{\sqrt{1+J^2}}-2\,\sqrt{(1+h)\ln
2}\right)
\end{eqnarray}
It's correct, when:
\begin{equation}
\label{XM}
\sqrt{(1+J^2)(1+h)\ln 2}\,>\,J_0^1+J\sqrt{h\ln 2}
\end{equation}
\begin{equation}
\label{XL}
\sqrt{(1+J^2)(1+h\ln 2}\,>\,J_0^2+\sqrt{h\ln 2}
\end{equation}
If the inequality (\ref{XM}) or (\ref{XL}) is violated, we have:
\begin{eqnarray}
\label{XY}
&&1-a\exp\left\{(-J_0^1-\sqrt{h\ln2})^2N\right\}
\theta(-J_0^1+2\sqrt{h\ln2})-\nonumber \\
&&-b\exp\left\{ -1(J^1)^2+h\ln2
N\right\}\theta\left[J_0^1-2\sqrt{h\ln2}\right]
\end{eqnarray}
for
\begin{equation}
\label{XP}
\left\{
\begin{array}{l} J_0^1+J\sqrt{h\ln2}\,>\,\sqrt{(1+J^2)(1+h)\ln2}\\
J_0^1+J\sqrt{h\ln2}\,>\,J_0^2+\sqrt{h\ln2}\\
\end{array}\right.
\end{equation}
or
\begin{eqnarray}
\label{XO}
&&1-b\exp\left\{-(\frac{J_0^2}{J})^2+h\ln2 N\right\}
\theta\left[\frac{J_0^2}{J}-2\sqrt{h\ln2})\right]\nonumber \\
&&-a\exp\left\{-(\frac{J_0^2}{J}-\sqrt{h\ln2})^2 N\right\}
\theta\left[-\frac{J_0^2}{J}+2\sqrt{h\ln2}\right]
\end{eqnarray}
for
\begin{equation}
\label{XW}\left\{
\begin{array}{l} J_0^2+\sqrt{h\ln2}\,>\,\sqrt{(1+J^2)(1+h)\ln2}\\
J_0^2+J\sqrt{h\ln2}\,>\,J_0^1+J\sqrt{h\ln2}\\
\end{array}\right.
\end{equation}

Similarly we can consider the situation for $J^2>h$.

We see (come to the conclusion) that for $T=0$ there are many
subphases, where there are different modes (regimes) of finite
size effects.\\
Perhaps it will be possible to apply Derrida's model generalizations
for the more complicated problems of information theory like broadcast
channels.\\
{\it Acknowledgments}
We thank Ye. Harutunyan for discussions.\\
This work was supported by  German ministry of Science and Technology
Grant 211-5231.


\begin{thebibliography}{99}



\bibitem{BD} B.Derrida, Phys.Rev.Lett.{\bf 45} (1980) 79.
\bibitem{ba} D.Saakian, JETP Lett.{\bf 61} (1995) n.8.
\bibitem{DG} D.Gross, M.Mezard, Nucl.Phys.{\bf B240} (1984) 43.
\bibitem{NS} N.Soarlas, Nature {\bf 239} (1989) 693.
\bibitem{DS} D.Saakian, JETP Lett.{\bf v.55} (1992) 798.
\bibitem{RD} D.Rujan, Phys.Rev.Lett. {\bf 701} (1993) 2968.
\bibitem{NH} H.Nishimori, Physica,{\bf A204} (1994) 1.
\bibitem{AAS} I.Csiszar, J.Korner, Information Theory, 1995
\end{thebibliography}
\end{document}